# SARG: THE HIGH RESOLUTION SPECTROGRAPH

R.G. Gratton[1], G. Bonanno[2], P. Bruno[2], A. Cali, R.U. Claudi[1], R. Cosentino[2,3], S. Desidera[1], G.Farisato[1], G. Martorana[1], M. Rebeschini[1], S. Scuderi[2], M.C.Timpanaro[2]

[1]*Osservatorio Astronomico di Padova*
[2]*Osservatorio Astrofisieo, Catania*
[3]*Centro Galileo Galilei TNG, Santa Cruz de La Palma (E)*
[4]*Dipartimento di Astronomia, Universitià degli Studi, Padova*

**Abstract**

SARG is the high resolution spectrograph of TNG. It has been in operation since late spring 2000. SARG is a cross dispersed echelle spectrograph; it offers both single object and long slit (up to 26 arcsec) observing modes covering a spectral range from $\lambda=0.37$ up to 1 µm, with resolution ranging from R=29,000 up to R=164,000. Cross dispersion is provided by means of a selection of four grisms; interference filters may be used for the long slit mode (up to 26 arcsec). A dioptric camera images the cross dispersed spectra onto a mosaic of two 2048x4096 EEV CCDs (pixel size: 13.5 µm) allowing complete spectral coverage at all resolving power for Å <0.8 µm. An iodine-absorbing cell allows to obtain high precision radial velocities. A Distributed Active Temperature Control System (DATCS) maintains constant the temperature of all spectrograph components at a preset value. Early results show that SARG works according to original specifications in terms of wavelength coverage, efficiency (measured peak efficiency is about 13%), resolution (maximum resolution R~164,000 using a 0.3 arcsec slit, R~144,000 using an image slicer), and stability (preliminary estimates of the radial velocity accuracy is ~5 m/s using the iodine cell and ±150 m/s without the cell).

## I. Introduction

SARG is the high resolution optical spectrograph for the Italian Galileo National Telescope (TNG). SARG was designed as a multi-purpose instrument, in order to satisfy the scientific needs of a rather wide community, working on a variety of themes, ranging from stellar abundance analysis or extended objects, to line profile studies and accurate radial velocity measurements. However, emphasis in the design was given to have a high resolution, stable instrument, while not compromising efficiency; such an instrument is particularly good for precise radial velocity programs, such as planets search and astero-seismology.

Instrument specifications included a high spectral resolution (maximum about R~150,000), high efficiency (peak >10% including telescope and detector), rather large spectral coverage in a single shot (two shots allowing to cover the entire accessible range from 370 to 900 nm; high sensitivity in the UV could not be obtained due to budget limitations), high stability (long term stability of 5 m/s, and possibly 1 m/s over a single night). While meeting all these specifications required a complex and sophisticated instrument, design criteria also included simplicity of use and maintenance, and respect for the TNG standards and environment. The project (approved in February 1996) was scheduled as a four year project, and within a typical budget for a 4 m telescope class instrument. To realize such a complex instrument within the allowed budget, we had to take responsibility for the optical design and integration of the instrument. Approximately following the original schedule, SARG first spectra were obtained in the evening of June 9, 2000; SARG was offered to all the TNG community from January 2001, slightly less than five years after its approval. As we will show SARG performances are within the original specifications.

SARG was built as a collaboration of the astronomical observatories of Padova (optics, mechanics, thermal control, integration, science verification, project management), Catania (controls, software, detectors), Trieste and Palermo (management and science verification). The iodine cell was realized as the doctoral dissertation program of S. Desidera (Dept. of Astronomy, Univ. Padova), with the collaboration of G. Favero (Dept. of Chemistry, Univ. Padova) and A. Cali (Catania Observatory).

## 2. Technical characteristics

The general layout of SARG is shown in Figures 1 and 2. SARG is permanently mounted on an optical table rigidly attached to the fork of the TNG, about 1.4 m below the Nasmyth B focus (see Figure 1). Light reaches the spectrograph through an optical train which includes three lenses, and a folding mirror (FMI) which redirects light exiting horizontally along the elevation axis downward to the spectrograph location. L1 and FMI are mechanically located within DoLoRes which permanently occupies the Nasmyth B location (see Fig. 1).

Note that due to SARG location, far from the Nasmyth focus, the TNG rotator adapter must be kept fixed while using SARG. Field derotation is provided by an optical derotator, and the standard TNG guiding camera of the derotator cannot be used.

In analogy with several modern high resolution spectrographs, SARG has a white pupil collimator. This design, coupled with the use of an R4 echelle, of grism cross dispersing elements, and of a large field, quite long focal length dioptric camera exploiting a large size detector, composed of a mosaic of 2 2k x 4k CCDs, pixel 13.5 µm, allowed a very compact and simple mechanical design. Instrument size is about 2x 1 x 1m, small in view of the high spectral resolution achieved by the instrument when mounted on a 3.5 m telescope. This compact design also simplifies its thermal design: a special feature of SARG is its distributed active thermal control system (DATCS), which allows the spectrograph to have a nearly constant temperature of 19.5±1.6 C while temperature in the Nasmyth B room changes from 0 to 20 C.

Table 1. SARG Grisms

|  | BLUE | GREEN | YELLOW | RED |
|---|---|---|---|---|
| Spectral Range (nm) | 369-518 | 419-567 | 462-795 | 502-1020 |
| Peack efficency (%) | 9.0 | 10 | 13.2 | 11.5 |
| Min. separation (arcsec) | 11.3 | 13.9 | 8.1 | 6.0 |

The preslit optics (which include a collimated part of the optical path) and the grism cross disperser also add flexibility to SARG, allowing a variety of optical modes: multiorder short slit observations with a large spectral coverage; 26 arcsec single order long slit observations (an optical derotator allows to have an arbitrary fixed orientation of the slit projected on the sky); multiorder image slicer observations at very high spectral resolution, using a Diego modified Bowen - Walraven image slicer (Diego 1994); high precision radial velocity observations using an I2 absorbing cell. Finally, a polarization analyzer will be mounted in front of the spectrograph in March 2001.

**Table 2. SARG Slits**

| Aperture | Resolution | Size (mm) | Size (arcsec) | Projected size (pixels) | Note |
|---|---|---|---|---|---|
| #1 | 29,000 | 0.30x1.50 | 1.60x8.0 | 9.8x49.1 | |
| #2 | 57,000 | 0.15x1.00 | 0.80x5.3 | 4.9x32.5 | |
| #3 | 86,000 | 0.10x1.00 | 0.53x5.3 | 3.2x32.5 | |
| #4 | 164,000 | 0.05x1.00 | 0.27x5.3 | 1.6x32.5 | |
| #5 | 43,000 | 0.20x5.00 | 1.07x26.7 | 6.6x 162.5 | |
| #6 | 115,000 | 0.075x5.00 | 0.40x26.7 | 2.2x 162.5 | |
| #7 | 144,000 | 0.30x0.30 | 1.60x1.6 | 2.0x68.6 | Image slicer |
| Pinhole | | 0.05x0.05 | | | Wavefront reference Mirror |

The main characteristics of the cross disperser and slits selection are summarized respectively in Table 1 and Table 2. For more information about SARG, see Gratton et al. 2001 and the SARG WEB Page at the TNG site: www.tng.iac.es

SARG uses a single, quite long (485 mm focal length, providing a scale of 0.16 arcsec/pixel when used at SARG), large field (8.5 degrees) dioptric camera constructed by SESO on our design.

Optical quality is very good, yielding an expected maximum FWHM of the instrumental profile due to optics of 13 µm (=1 pixel) over the whole field. Indeed we see only limited degradation of the optical quality of the system even at the highest spectrograph resolution (164,000). Also the efficiency of the camera is very high (on average, 90% over the useful spectral range 370-900 nm, with a minimum value of 84% at the shortest wavelength).

The detector is a mosaic of 2 thinned, back illuminated EEV 2k x 4k CCDs, pixel 13.5 µm providing a scale of 0.163 arcsec/pixel. The CCD's columns are aligned with echelle resolution minimizing the number of orders affected by the 400 µm (equivalent to about 30 pixels) gap between the two chips. Furthermore, the portion of the spectrum falling into this gap may be observed using a different grism. The measured RON is 7.5 e$^-$, and the conversion factor is about 1.55 e$^-$/ADU, at a read-out rate of ~60 Kpixels/sec. A variety of readout modes are implemented, allowing to bin the CCD (1x1, 1x2, 1x4, 2x1, 2x2, 2x4, 4x1, 4x2, 4x4). All these modes may be software selected through the SARG User Interface.

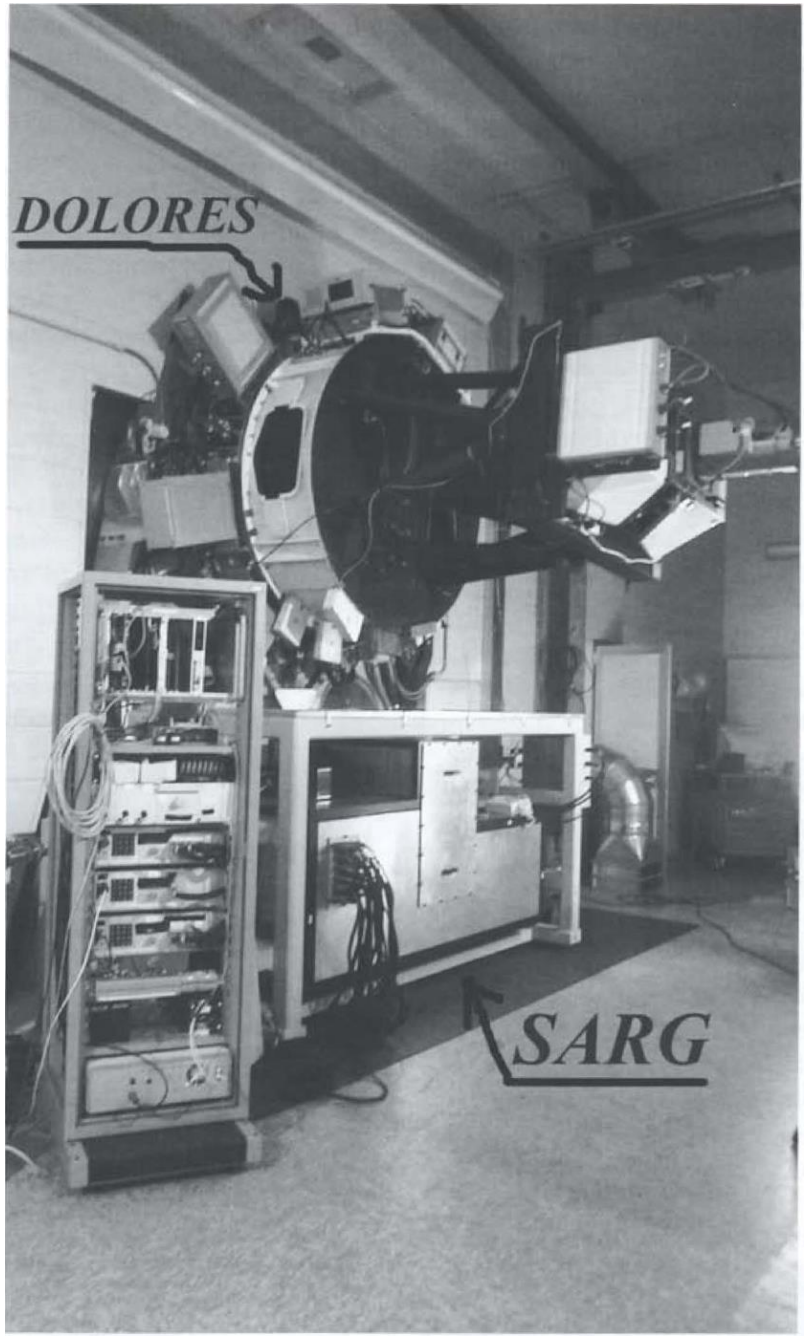

Fig. 1 - SARG layout in the TNG Nasmyth room B. On the foreground, the SARG electronic rack. Behind it, the SARG, rigidly mounted on the telescope fork. Light reaches the spectrograph from a folding mirror (in this picture hidden by the DOLORES structure). The rotator adapter is kept fixed when using SARG. Field derotation is achieved by means of an optical derotator. Guiding is done at the slit, using a cooled CCD.

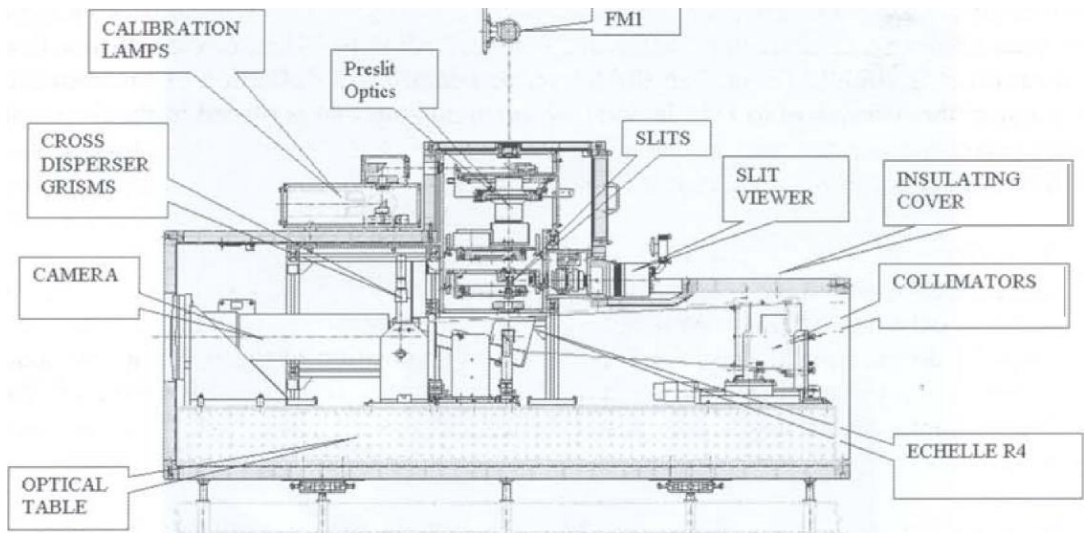

Fig. 2 - SARG layout. Light enters the spectrograph from top, after having being folded by a 45° degree mirror (FM 1). Note (in order along the optical path) the calibration lamp unit, the preslit optics(including the optical derotator and the preslit slide equipped with the absorbing cell), the slit wheel, the slit viewer, the spectrograph shutter, the white-pupil collimator, the R4 echelle, the wheel carrying the grisms used as cross dispersers, the dioptric camera, the optical table, and the insulating cover.

SARG is equipped with a set of auxiliary devices. This set includes:

- the **slit viewer**, based on a double-stage (air) Peltier cooled scientific camera equipped with a frame transfer, front illuminated CCD EEV0520 (770x1 152 pixels, 22.5 µm pixel size). The slit viewer has two modes of operation: direct imaging (using a Nikon objective Nikkor 105 mm f/2.5: with this mode the scale is 0.137 arcsec/pixel) and wavefront analysis (using a 22x22 Adaptive Optic Associated 0497-97-S-X Shack-Hartmann lenslet array: in this mode the scale is 0.149 arcsec/pixel; the spot layout is hexagonal, each spot being at ~22 pixels from the adjacent ones). This last mode allows to close the active optics loop (the reference beam is provided by a pinhole located on the slit wheel). These two modes may be selected by moving a commercial motorized slide. Due to the quite high dark current of the Slit Viewer camera, it is possible to point and guide on objects as faint as magnitude 16.5 (in typical observing conditions, with a bright near full moon). This currently sets the faintest limiting magnitude observable with SARG;

- the **Calibration Lamp Unit**. This may be used to feed the spectrograph with light provided by one lamp that may be software selected amongst a set of five (three quartz iodine halogen lamp for flat fielding; a Th lamp for wavelength calibration; and a Hg isotopic lamp for alignment purposes) by rotating a motorized stage carrying the Lamp Selection Mirror. Light from the lamp can be introduced into the optical path by means of a 45 degree folding flat mounted on the motorized Calibration Mirror Slide located in front of L2. The optics of the calibration lamp unit (realized using commercially available optics) was designed to efficiently and homogeneously illuminate an area of at least 5.4 mm diameter in the slit plane (corresponding to 24 arcsec diameter) with a beam that is slightly faster than the telescope beam. Final definition of the entrance pupil of the instrument to F/11 is done by the pupil stop that is placed in the image of the telescope secondary (nominal size 6 mm). Some vignetting (due to L3) occurs at distances larger than 12 arcsec from field center.

- **An iodine gas absorbing cell** ($I_2$, cell), that may be inserted in the optical path to imprint on the object spectrum an absorption spectrum of $I_2$ lines which can be used as a very precise and stable wavelength reference. The SARG $I_2$ cell is mounted on a specialty devoted position on the Preslit Slide. The position of the $I_2$ cell allows also taking a lamp flat through the cell. This is useful for the calibration of the spectrograph, and for engineering tests of absolute stability and stray light in the cores of saturated $I_2$ lines.

There are 9 motorized wheels and tables that allow remote control of all the spectrograph functions. Five of them are realized by means of commercial (OWIS) tables; the remaining four are annular rotating tables realized by CINEL. Figure 3 illustrates the SARG motorized parts

Fig. 3 - SARG motorized components

A sketch of the overall architecture of SARG controls is given in Figure 4. It is a VME-based system; the VME-bus directly controls the Elettromare TNG standard CCD controller by means of a fiber optics link and TNG standard Atenix boards. Instrument functions are controlled by RS232 links to stand-atone controllers (three OWIS motor-controllers, a National board commanding the lamps, and a Lakeshore temperature monitor: see Figure 5). All these elements are located on a single rack, a couple of meters far from the spectrograph.

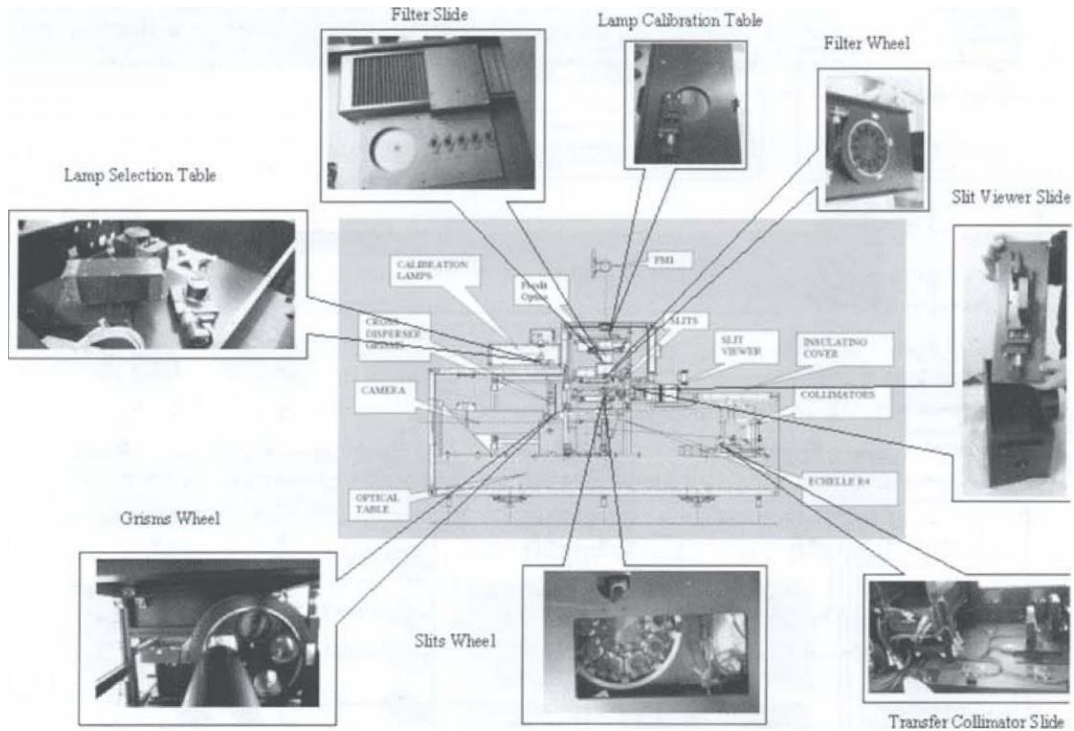

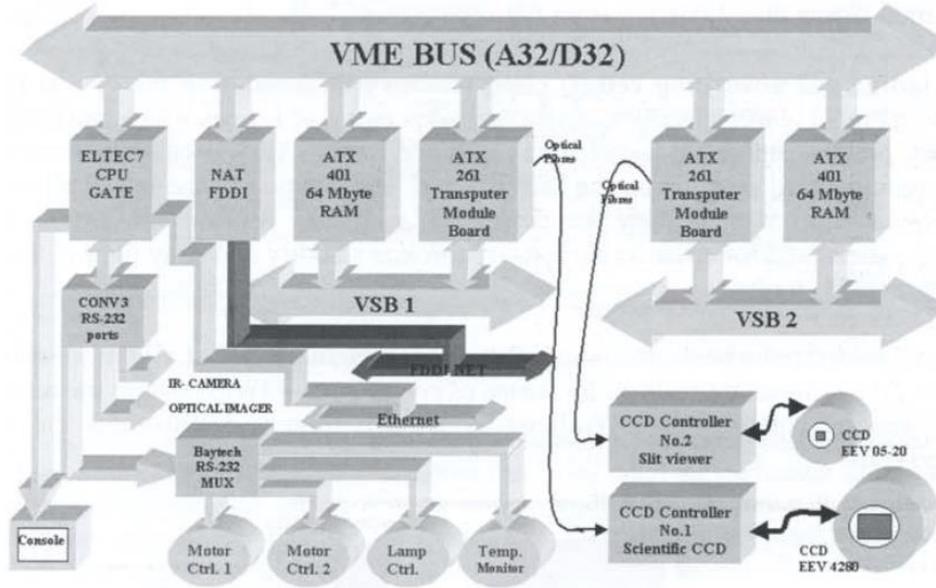

**Fig. 4 - Architecture of SARG controls**

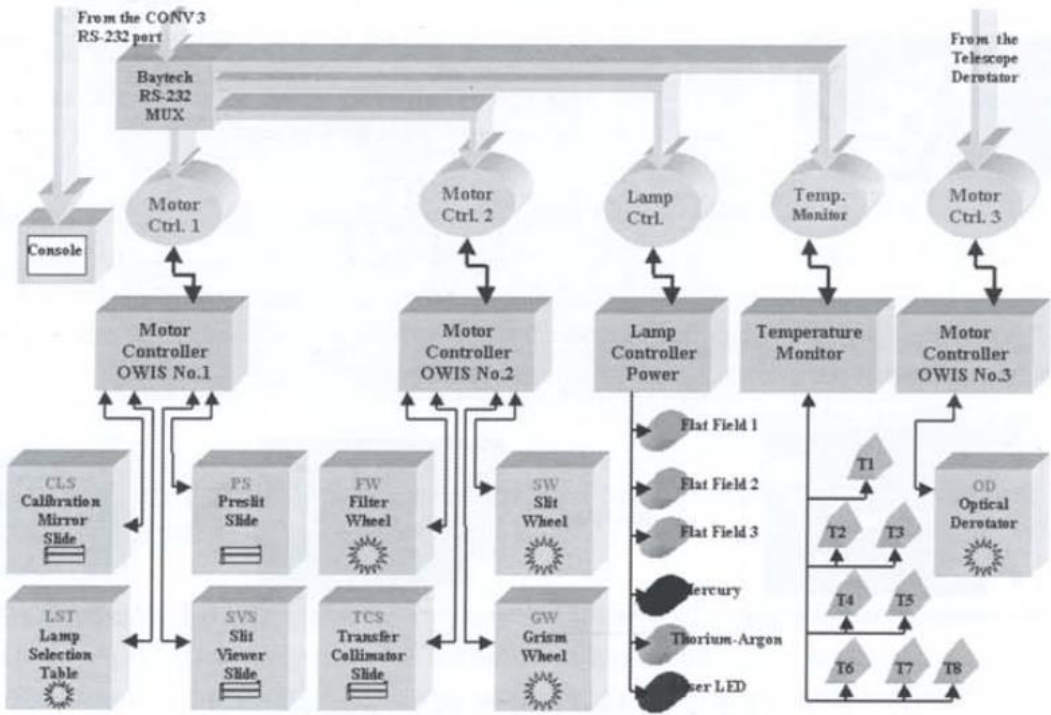

**Fig. 5 – Scheme of low level SARG controls**

A high level graphic user interface (GUI) is available in order to control all SARG functions. In Figure 6a the main window of the GUI is shown, while in Figure 6b a sketch of the logical structure of the control software User Interface is given.

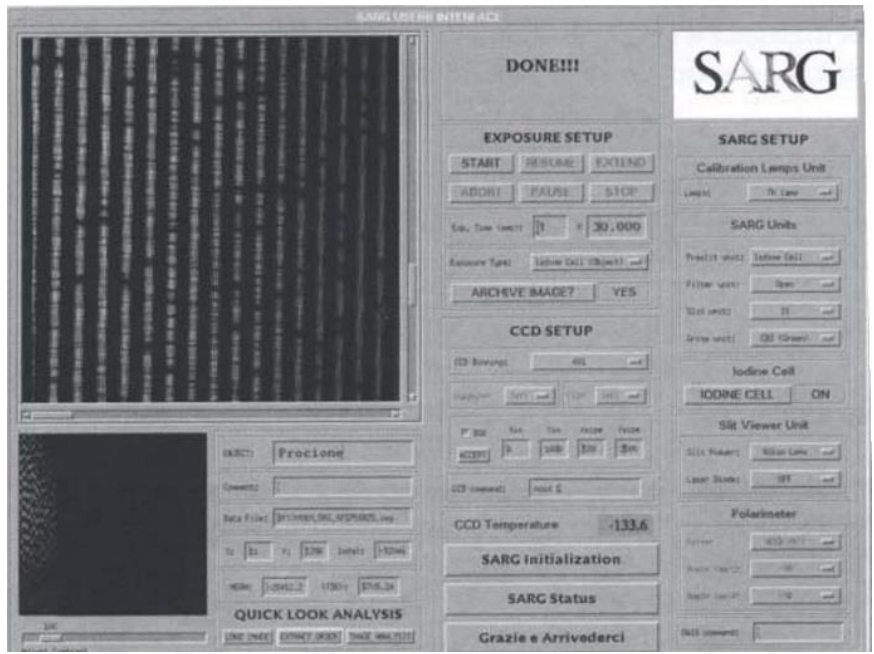

a)

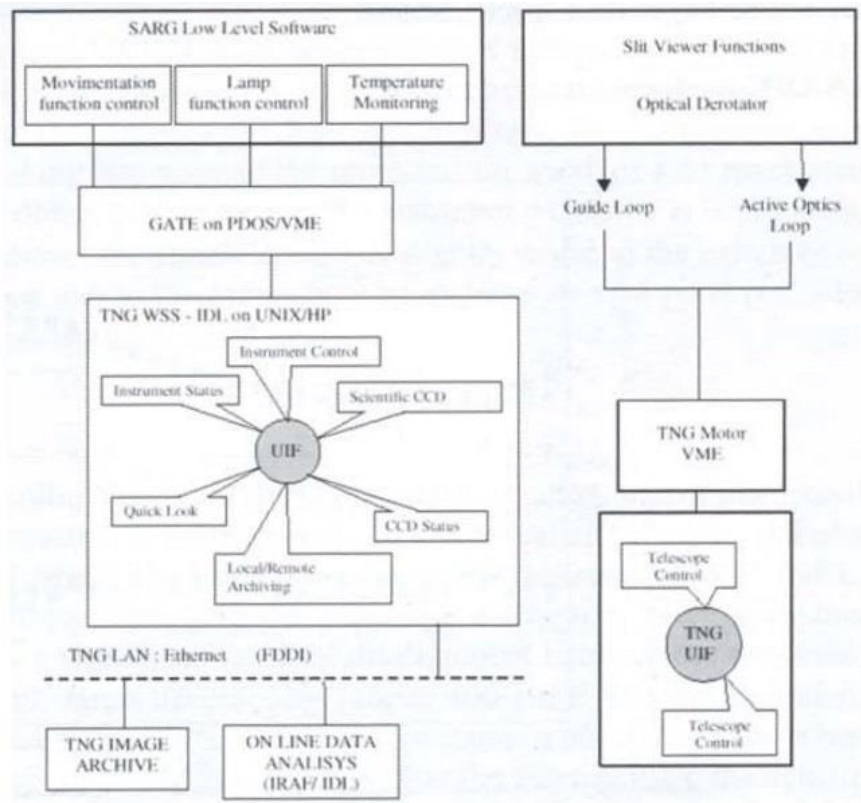

b)

*Fig.6 – Panel a: The main window of SARG user interface.*
    *Panel b:* Structure of SARG software

## 3. SARG Distributed Active Thermal Control System (DATCS)

Temperature within the SARG enclosure is kept approximately constant by means of an active thermal control system. The design of this system was done using a finite element computer model, tested against suitable laboratory experiments (see Gratton et al. 2001 for details). The SARG thermal specifications were set as follows: (i) Long-term stability: <0.5 C over 10 yr; (ii) Thermal dishomogeneity: <1 C; (iii) Rate of temperature changes: <1 C/hr.

To achieve specifications, DATCS architecture consists of two main components:

- 30 MINCO CT 198 HEATERSTAT sensorless temperature controllers coupled with thermofoil resistances;
- 8 DT-470 silicon diode sensors controlled by a Lakeshore 208 temperature Monitor.

We expected DATCS to keep temperatures within SARG at 19.7±0.7 C while the temperature in the Nasmyth room changes from 0 to 19 C, the range given in TNG specifications. The whole optomechanical design was based on this specifications: for instance, based on these values we do not expect that any refocusing of the spectrograph camera is required (actually the spectrograph might be refocused by moving the transfer collimator; however we foresee this possibility only when the image slicer is used). Strict procedures will be enforced to avoid that during spectrograph heating temperature gradients will not be larger than specifications.

## 4. SARG performances

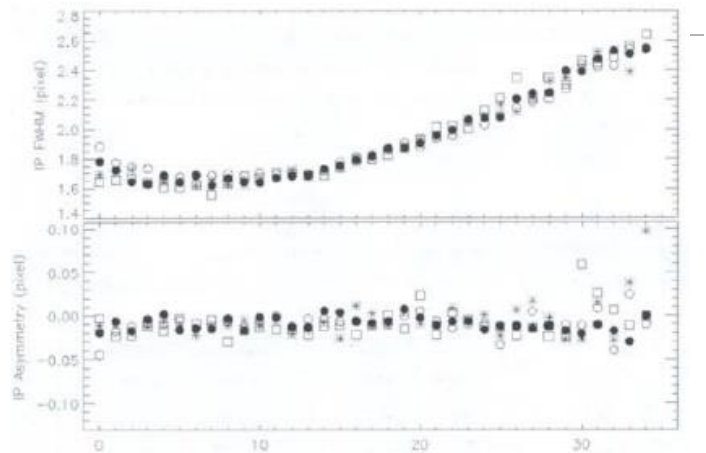

Figure 7. Run of the FWHM and of the asymmetry (defined as the difference between the left HWHM and the right HWHM) of instrumental profiles derived from a B star + Iodine for different spectral orders (120: empty circles; 115: filled circles: 111: asterisks; 106: open squares). Each chunk represents a portion of 101 pixels along dispersion.

### *4.1 Optical quality and resolution*

The most critical tests of the SARG optical quality are those provided by the instrumental profiles (IPs) derived from study of the iodine lines. The software used to extract high precision radial velocities (AUSTRAL) allows a deconvolution (using a maximum entropy method) of the observed iodine cell spectrum, using as a template a very high resolution (R~750,000), high S/N (≥500) spectrum obtained with the Kitt Peak Fourier Transform Spectrograph (FTS).

Figure 7 shows the run of the FWHM and of the asymmetry (defined as the difference between the left HWHM and the right HWHM) of instrumental profiles derived from a B star + Iodine for different spectral orders, obtained with the narrowest slit (width: 50 µm that is a projected slit width of 1.63 pixels at order center). The width of the instrumental profile shows a trend as a function of the position along the order, changing on average by 40% level (from 1.6 to 2.4 pixels). Such a difference is due to the anamorphic magnification introduced by the R4 grating, that is changing from 0.93 to 1.35 at the extremes of the orders visualized on the CCD (the projected slit width changes from 1.52 to 2.20 pixels: note that orders are not exactly centered on the CCD, the center being at chunk ~10 on Figure 7). The slight curvature of the run of the FWHM (causing additional image broadening at blue and red edges of the order) is caused by a slight defocusing because the focal surface is not exactly planar. The instrumental profile shows a nearly symmetric shape. Differences between left and right HWHM are below 3%.

From these tests we derive a typical FWHM at the order center of 1.68 pixels. Including the small contribution of Kitt Peak FTS to the intrinsic iodine spectrum, the resulting FWHM of SARG instrumental profile at the center of the order is 1.76 pixels. This corresponds to a resolving power of 164,000. Taking into account the projected slit width of 1.63 pixels, we derive that at order centers the optics contribution to the instrumental profile is 0.66 pixels, that is ~9 µm. As mentioned above, the optical quality is slightly worse at the extremes of an order, where the contribution due to the optics may be as large as ~1.4 pixel (i.e. ~19 µm: slightly worse than the expected value of 13 µm).

### *4.2 Efficiency*

Owing to the small size of the collimated beam in the preslit optics, alignment of the optical and mechanical axes of the derotator is very critical. This was realized directly at Fisba Optiks, working the derotator wheel (an annular rotating table constructed by CINEL, providing a very stable rotation axis, with no flipping). However, even after this painstaking procedure, there is a significant residual disalignment between the two axes, with a small tilting but a quite large decentering (about 400 µm). The misalignment between the optical and mechanical axes of SARG derotator causes a pupil migration when changing derotator position. This causes some vignetting (on the filter unit, on the echelle and on the cross disperser). The efficiency of the spectrograph for different derotator positions was measured on September 13, 2000 by observing without a slit (before the image slicer was mounted) the star ξ Cas (see Figure 8). A maximum of efficiency (at 5200000 counts) is clearly identified. Visual inspection of pupil footprints within the spectrograph showed that there is no vignetting when the derotator is in this position.

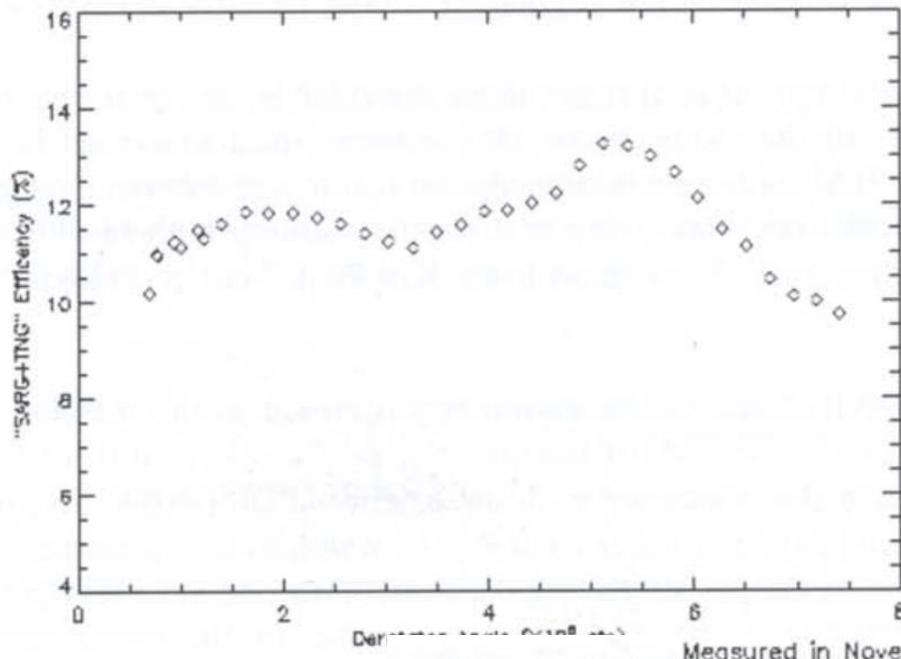

**Figure 8. Efficiency of TNG+SARG as a function of the position angle of the optical**

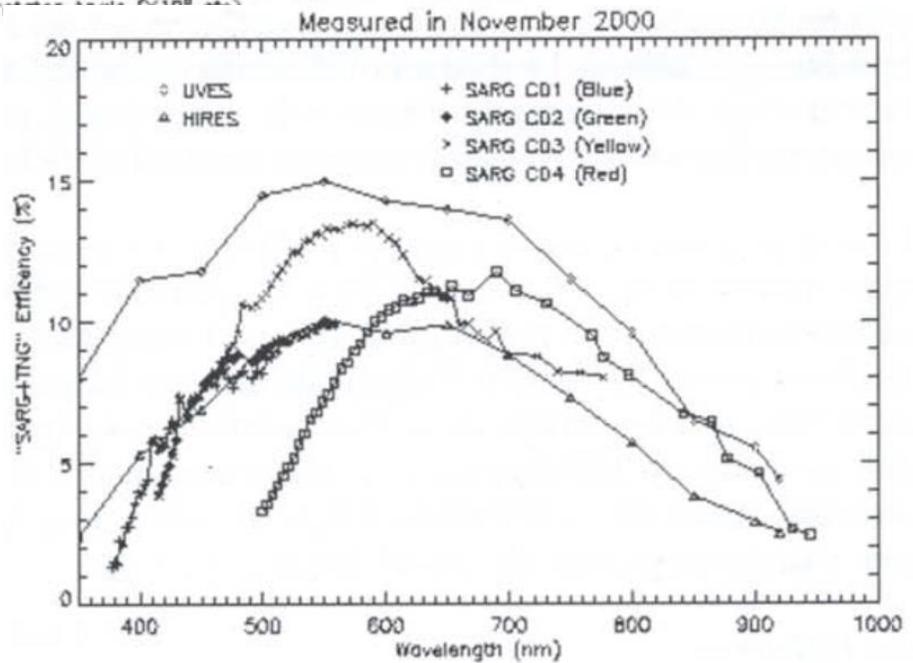

**Figure 9. TNG+SARG efficiency (without slit) for all available grisms, measured observing the spectrophotometric standard 58 Aql on Nov. 5, 2000. For comparison, similar measurements for UVES at Kueyen (VLT Unit 2) and HIRES at Keck are also shown**

The efficiency of the spectrograph in various configurations (but with the derotator always in its optimal position) was measured on a photometric night (November 5, 2000) observing the spectrophotometric standard 58 Aql (HR 7596), again without any slit. The efficiency resulted higher than measured in September, because the TNG mirrors had been cleaned a few days before (note that the TNG primary mirror was last aluminized in September 1999, about one year before the present tests). The measured efficiency of TNG+SARG using the four grisms is given in Figure 9. For comparison, similar results for UVES at Kueyen (VLT Unit 2), and HIRES at Keck are also given. At peak wavelength, SARG+TNG efficiency is only slightly lower than that measured for Kueyen+UVES (however, TNG mirror coatings are probably somewhat less efficient than those of Kueyen), and well above the values obtained for Keck+HIRES.

## *4.3 Thermal behavior and stability*

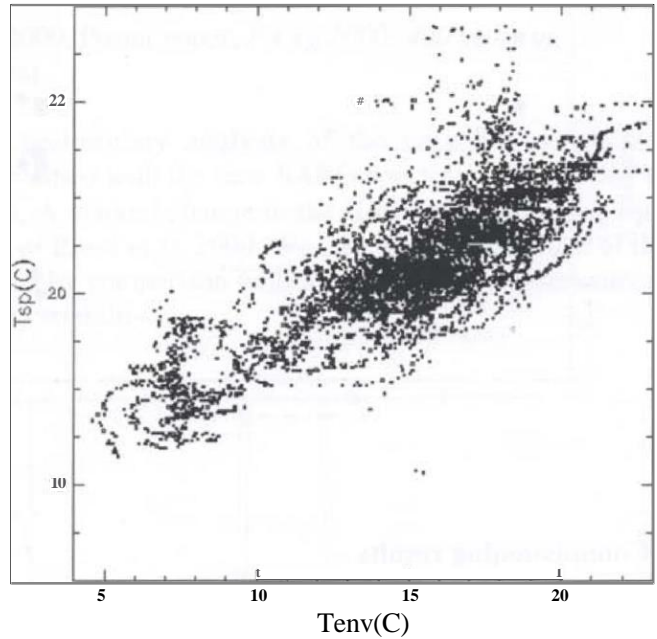

Figure 10. SARG temperature $T_{sp}$ as a function of temperature in the TNG Nasmyth B room $T_{env}$, during the period July-November 2000

As a test of the thermal behavior of SARG, we show in Figure 10 the run of the SARG temperature $T_{sp}$ as a function of temperature in the TNG Nasmyth B room $T_{env}$ during the period July-November 2000 (note that different sensors within the spectrograph yield the same results within 0.2 C). On the whole, temperature within the spectrograph changes by ~0.16 C for a change of 1 degree in the temperature in the Nasmyth room. Since the expected temperature extremes in the Nasmyth room are 0 to 20 C, we expect a total temperature excursion of ±1.6 C around a mean value of about 19.5 C. The sensitivity of SARG temperature to the external one is about twice the value expected on the basis of our models.

## *4.4 Radial velocity precision*

High precision radial velocities are one of the main targets of SARG. The whole potentiality of SARG in this field is still to be established. A first definition of the medium term (one month stability) of SARG radial velocities using the Iodine Cell is given by Figure 11, which gives early results obtained from a preliminary analysis of 14 spectra of $\tau$ Cet, a star believed to have constant radial velocity. They were obtained using a single order, out of the about 25 covering the useful spectral regions with lines of the iodine cell. The r.m.s. of these measures is 7.8 m/s. We expect significant improvements from the use of the full useful spectral range.

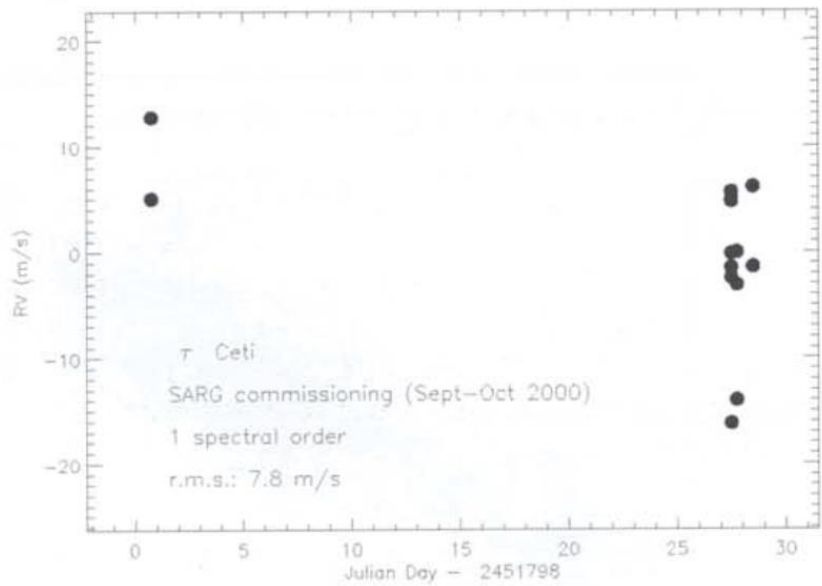

Figure 11. Preliminary radial velocities for the standard star τ Cet obtained from a single order. Full reduction will make use of the whole spectral region including lines of the iodine cell (about 25 orders)

## 5. Commissioning results

- **Abundances in old open clusters: NGC6791 and NGC 6819** (Bragaglia et al. 2001

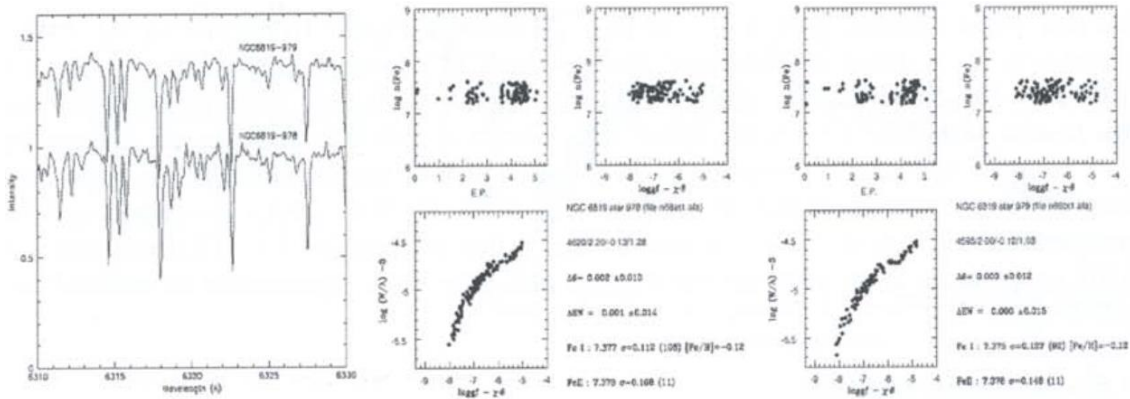

Figure 12: Representative graphs showing results obtained for the program on old open clusters. Note the extremely good definition of the curve obtained for stars with m>13.

We present an analysis of high dispersion spectra (R~40000) of three red clump stars in the old open cluster NGC 6819. The spectra were obtained with SARG. They were analyzed using both equivalent widths measured with an automatic procedure, and comparisons with synthetic spectra. NGC 6819 is found to be slightly metal rich ([Fe/H]=+0.09 ± 0.03, internal error); there are no previous high resolution studies to compare with. Most element to element abundance ratios are close to solar; we find a slight excess of Si, and a significant Na overabundance. Our spectra can also be used to derive the interstellar reddening towards the cluster, by comparing the observed colors with those expected from line excitation: we derive E(B-V)=0.14 ±0.04, in agreement with the most recent estimate for this cluster.

- **P Cyg and AG Dra** (Rossi et al. 2000, Poster paper, *P Cyg 2000: 400 years of progress, PASP Conf Ser.,* in press)

We present here the results of a preliminary analysis of the very high resolution spectrograms (R=86000) of P Cyg obtained with the new SARG spectrograph attached to the Italian National Telescope (TNG). A marked change in the Hα profile is noticed with respect to the May 1996 observations of Rossi et al. 2000. We also analyze the effect of the telluric lines on the spectral appearance by comparison with lower resolution observations, and find that it can simulate weak line variability.

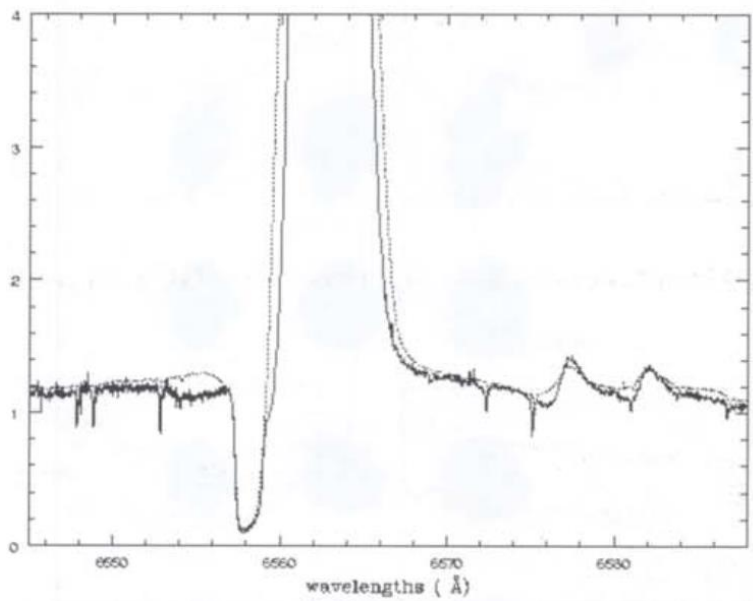

Figure 13: Hα profile variation at very high resolution, The June 2000 SARG spectrum (solid line) is compared with that of May 1996 (from Rossi et al. 2000, dotted line).

- **StHα 190** (Munari et al, 2001, A&A submitted)

A highly and rapidly variable bipolar mass outflow from StHα 190 has been discovered, the first time in a yellow symbiotic star. Permitted emission lines are flanked by symmetrical jet features and multi-component P-Cyg profiles, with velocities up to 300 km s$^{-1}$. Given the high orbital inclination of the binary, if the jets leave the system nearly perpendicular to the orbital plane, the de-projected velocity equals or exceeds the escape velocity (1000 km s$^{-1}$). StHα 190 looks quite peculiar in many other respects: the hot component is an 0-type sub - dwarf without an accretion disk or a veiling nebular continuum and the cool component is a G7 III star rotating at a spectacular 105 km S$^{-1}$ largely unseen in field G giants.

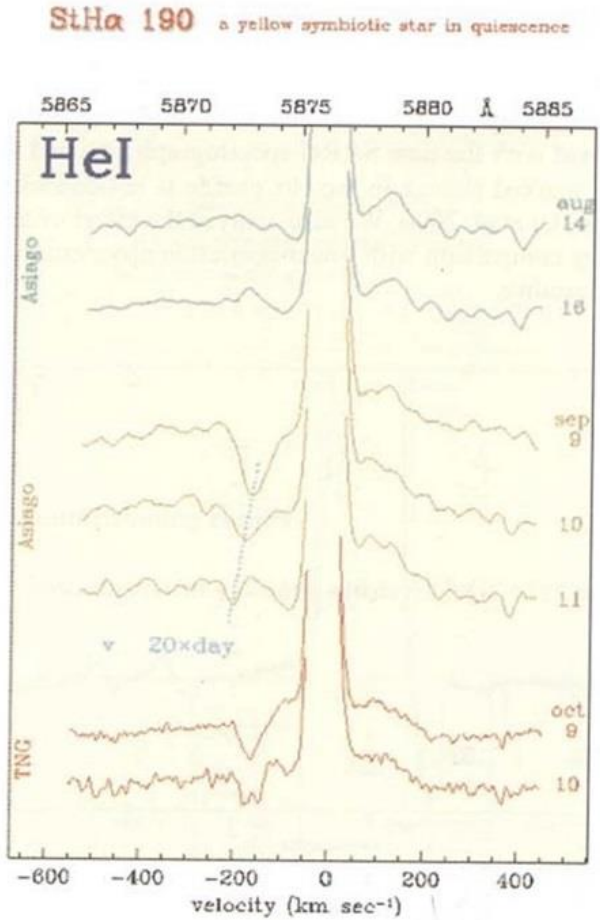

Figure 14: Evolution of the He I emission line profiles over a 4 month period (see dates on the right); SARG spectra are named TNG.

- **Tomography of selected planetary nebulae:** (Sabbadin, Ragazzoni, reduction in progress)

Figure 15 (next page): <u>Left panel:</u> Representative emission line structures in NGC6572 at P.A.=0º (SARG + red grism; 5 min. exposure; spectral resolution 115,000). In this reproduction the observed intensities are enhanced (by the factor indicated in parenthesis) to make each line comparable with $H_\alpha$. Note the blurred appearance in $H_\alpha$ (mainly due to thermal motions) and the large stratification effects present in the nebula: low ionization emissions, such as [OI] and [NII], occur in the external, faster layers, whereas the high excitation ones, [ArIII] and [0III], are located in the internal, slower expanding regions, <u>Right panel</u> : Top: Main [OIII] λ 5007 Å, $H_\alpha$, and [NII] λ 6584 Å position - velocity structures observed at four position angle of NGC6572 (instrument configuration as before). The intensities of [0III] and [NII] have been scaled to $H_\alpha$. Bottom Same, but at higher contrast, to show the faintest [OIII], Hα and [NII] structures present in the 5 min SARG echellograms (spectral resolution 115,000) of NGC6572.

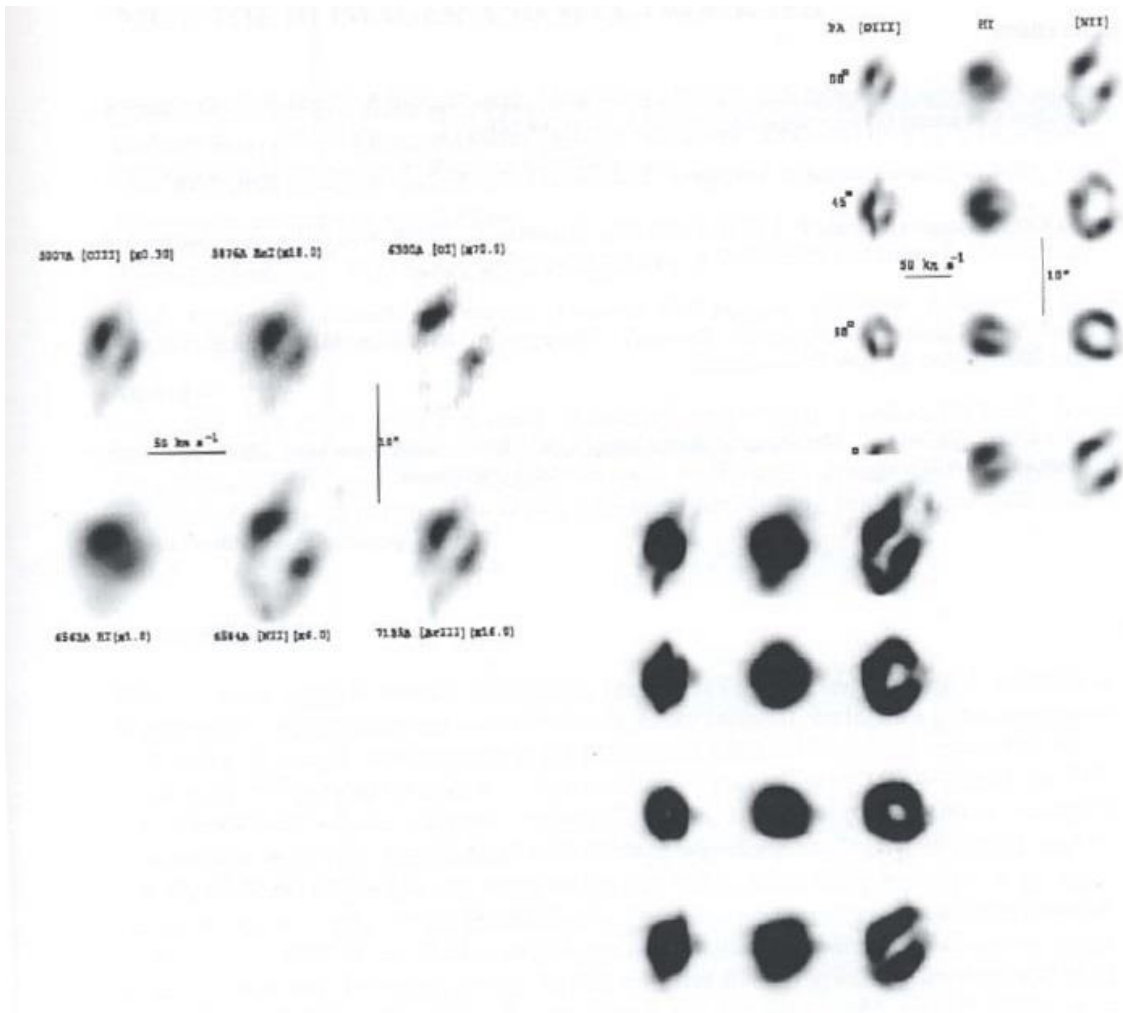

Figure 15.

Other works are already in progress:

- Doppler imaging of HR 1099 (Rodonò, Marino et al., reduction in progress);

- Search for planets in visual binaries (SARG Team, survey started; reduction in progress);

- Magnetic stars (S. Catalano, Leone et al., reduction in progress);

- Solar type stars in Hyades and Pleiades (S. Catalano, Micela, Pallavicini, to be completed)